\documentclass[twocolumn,times,astrosymb]{aastex631}

\newcommand{\mPO}{P(O)}  
\newcommand{\PO}{$\mPO$}

\newcommand{\mPU}{P(U)}  
\newcommand{\PU}{$\mPU$}

\newcommand{\mPOx}{P(O|x)}  
\newcommand{\POx}{$\mPOx$}

\newcommand{\mPUx}{P(U|x)}  
\newcommand{\PUx}{$\mPUx$}

\begin{document}

\title{Effects of localisation precision on identified fast radio burst host galaxy magnitudes}

\author[0000-0002-6437-6176]{Clancy~W.~James}
\altaffiliation{E-mail: clancy.james@curtin.edu.au}
\affiliation{International Centre for Radio Astronomy Research, Curtin University, Bentley, WA 6102, Australia}

\author[0000-0002-7738-6875]{J.~Xavier~Prochaska}
\affiliation{Department of Astronomy and Astrophysics, University of California, Santa Cruz, CA 95064, USA}
\affiliation{Kavli Institute for the Physics and Mathematics of the Universe, 5-1-5 Kashiwanoha, Kashiwa 277-8583, Japan}
\affiliation{Division of Science, National Astronomical Observatory of Japan, 2-21-1 Osawa, Mitaka, Tokyo 181-8588, Japan}

\author[0000-0002-2864-4110]{Apurba Bera}
\affiliation{International Centre for Radio Astronomy Research, Curtin University, Bentley, WA 6102, Australia}

\begin{abstract}

We study the potential bias in the identification of fast radio burst (FRB) host galaxies due to radio localisation uncertainty. Using a sample of FRBs localised to typically $0.5^{''}$ by the Australian Square Kilometre Array Pathfinder (ASKAP), we artificially increase the localisation uncertainty up to $10^{''}$, and re-run the Probabalistic Association of Transients to their Hosts (PATH) algorithm to determine the most likely host galaxy. We do not find evidence of a significant change in identified hosts until the localisation precision is worsened to $2^{''}$ or greater.
\end{abstract}


\section{Introduction} \label{sec:intro}

Since their discovery in 2007 \citep{Lorimer2007}, the nature of the progenitor(s) of fast radio bursts (FRBs) has been of great interest.
Enabled by precise radio localisations of FRBs \citep{Chatterjee2017,Bannister2019}, various authors have analysed increasingly large samples of FRB host galaxies \citep{Bhandari2020a,2020ApJ...899L...6L,Heintz2020,Bhandari+22,Gordon+2023}. Their results have yielded broadly consistent conclusions with increasing precision: that the host galaxies of apparently once-off FRBs are inconsistent with the high-star-formation rate dwarf galaxies expected from superluminous supernovae (SLSNe) and long-duration gamma-ray bursts (LGRBs) \citep{2006Natur.441..463F}, while repeating FRBs tend to be found in such galaxies
\citep{Tendulkar2017}; 
FRBs rarely occur in ``red and dead''
galaxies which have ceased star-formation
\citep[e.g.][]{shah2025,te2025}, 
and do not trace total stellar mass \citep[e.g.][]{sph2020}; 
their mass-metallicity follows standard $0.07<z<0.7$  relations \citep{Heintz2020}; and their host populations are broadly consistent with those of core-collapse supernovae (CCSNe) and short gamma-ray bursts (sGRBs). 

Most recently, \citet{DSA2024SFR} expanded the sample with 30 host galaxies detected by the Deep Synoptic Array (DSA), confirming both the trend of FRBs tracing star-formation in high-mass galaxies, and a possible deficit of low-mass host galaxies compared to expectations from star-formation in the FRB host sample. In particular, the authors identify a break point of approximately $10^{10}\,M_* \,[M_\odot]$, above which FRB hosts do trace star formation, and they argue this is evidence for FRBs being associated with high metallicity environments.

In this work, we consider the degree to which the FRB's localisation causes a bias in identifying the FRB host galaxy in optical images from radio localisations, and hence, whether or not the observed deficit of dwarf hosts may be explained by such a bias.

\section{A bias in FRB host galaxies?}

Fast radio burst host galaxies are identified from optical follow-up observations in the direction of the radio signal, with an optical counterpart coincident with the radio localisation region being taken as evidence for that galaxy being the true host. This process produces a bias against low-luminosity FRB host galaxies in two ways. Firstly, such galaxies will in-general be dimmer, and hence less likely to be detected in optical images. Secondly, the prevalence of low-luminsoity galaxies makes the firm identification of any given one as the true host more difficult, even if it is detected in the image. The importance of very precise radio localisation in identifying low-luminosity FRB hosts was highlighted by \citet{2017ApJ...849..162E}, who noted that a localisation accuracy of 0.5'' would be required to identify a dwarf FRB host galaxy at $z \ge 0.1$ at 99\% confidence. Hence, dwarf FRB hosts will be less likely to be confidently identified, and hence less likely to be included in FRB host galaxy analyses, in a manner which is dependent upon the radio localisation accuracy.

The current standard method for quantifying the level of confidence in FRB host galaxy identification is the Probabalistic Association of Transients to their Hosts \citep[PATH;][]{PATH}. This Bayesian method applies a 
Jeffrey's prior \PO\ according to a galaxy's magnitude 
(typically the r-band magnitude, $m_r$); 
a prior $P(\theta/\phi)$ on the distribution of angular offsets $\theta$ of an FRB from the host centre relative to its half-light radius $\phi$; and the probability \PU\ of the true host being unseen in the image. PATH also accounts for the radio localisation precision, usually in the form of an elliptical Gaussian. The output is a posterior estimate, \POx, of each host galaxy candidate's probability of being the true host, and a probability \PUx\ of the true host remaining undetected.

A full determination of any possible bias in PATH would 
be complex, and depend on the assumed underlying distribution of FRB host galaxies. Here, we perform a simple test of this potential bias using FRBs detected by the CRAFT Incoherent Sum (`ICS') survey \citep{Shannon2024ICSsurvey}.

\begin{figure}
    \centering
    \includegraphics[width=\linewidth]{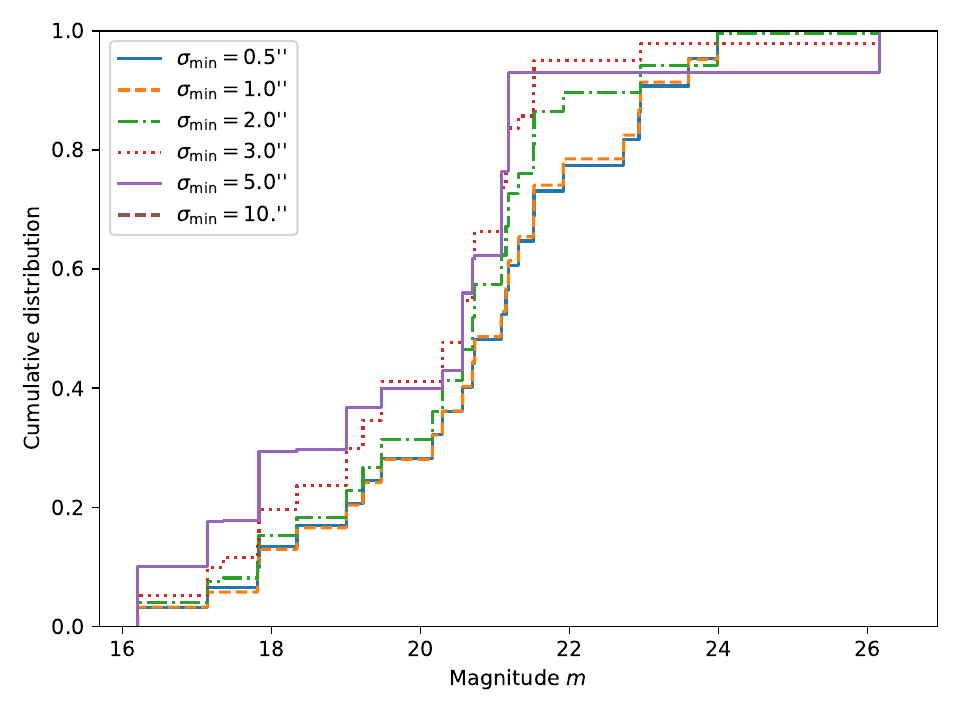}
    \caption{Cumulative distribution of magnitudes, $m_r$, for confidant host galaxy identifications from the CRAFT/ICS sample, for different relative localisation errors.
    }\label{fig:magnitude_distributions}
\end{figure}

\section{Simulation of the bias using the ASKAP/CRAFT ICS Sample}

The ASKAP ICS survey recently reported the detection of 43 FRBs \citep{Shannon2024ICSsurvey}. Of these, 37 had interferometric localisations with typical accuracy of $0.5^"$. A PATH analysis identified 30 host galaxies at $>95$\% confidence. 

We study the effect of localisation accuracy on this host galaxy sample by artificially degrading the localisation accuracy, and re-running the PATH analysis using the 
32~host galaxies with deep imaging (i.e.\ complete to an r-band magnitude of at least 25) currently provided in the
{\sc FRB} repository \citep{frb} with tables of 
PATH outputs. 
To do so, we increase the true localisation uncertainty, $\sigma_{\rm true}$, to a certain minimum, $\sigma_{\rm min}$, to generate a new localisation uncertainty $\sigma^\prime = {\rm max}(\sigma_{\rm true},\sigma_{\rm min})$. 
We then deviate the original position accordingly by $\sigma_{\rm extra} = \sqrt{\sigma^{\prime 2} - \sigma_{\rm true}^2}$. We repeat this randomisation 50 times for each FRB, place a cut of $\mPOx > 0.95$ on the output, and build up histograms of the resulting host magnitude distribution. For this analysis, we use a prior on the probability of the true host being unseen, $P(U)$, of $0.1$, comparable to that used in the work of \citet{DSA2024SFR}.

Figure~\ref{fig:magnitude_distributions} shows the cumulative distribution of firm host galaxy magnitudes for $\sigma_{\rm min}=0.5''-10''$. We find no effect for $\sigma_{\rm min}=1''$ and below, but that for $\sigma_{\rm min}=2''$ and above, the probability of identifying a low-luminosity ($m_r > 21.5$) host decreases, and the cumulative distribution of host magnitudes shifts downwards, to brighter galaxies. For $\sigma_{\rm min} = 5''$, very high-magnitude galaxies ($m_r < 19$) are favoured above those in the ($19 < m_r < 21$) range. When $\sigma_{\rm min}=10''$, no host galaxy is confidently identified.

\section{Conclusion}

We have tested the predictions of \citet{2017ApJ...849..162E} regarding the required accuracy to confidently identify FRB host galaxies, using a real sample of localised FRBs detected by ASKAP/CRAFT, and the PATH statistical framework.
We have shown that increased uncertainty in localisation accuracy begins to affect the implied FRB host galaxy magnitude distribution in the ASKAP ICS survey for localisation precisions $\ge 2^{''}$. We do not therefore expect there to be a significant localisation bias in the ASKAP and DSA FRB samples --- which have typical localistion accuracies of $\le 1''$ --- that are used in an analysis of FRB host galaxies. MeerTRAP coherent FRB detection data should be likewise useful \citep{Meertrap_2023_sample}. Note that this is not a comprehensive review of potential sources of bias in such analyses, and we do not exclude that other sources of bias may be present. For instance, the localisation accuracy of CHIME FRBs prior to the use of outriggers is several arcminutes \citep{CHIME_catalog1_2021}, and several FRBs detected by CHIME were used in the analysis of \citet{DSA2024SFR}.

\section*{Acknowledgements}

We acknowledge the traditional custodians of the land this research was conducted on, the Whadjuk (Perth region) Noongar people and pay our respects to elders past, present and emerging. CWJ acknowledges support by the Australian Government through the Australian Research Council's Discovery Projects funding scheme (project DP210102103).

\software{{\sc numpy 2.2.1} \citep{Numpy2011}, {\sc scipy 1.15.0} \citep{SciPy2019}, {\sc astropy 7.0.0}  \citep{2022ApJ...935..167A}, {\sc astropath} \citep{frb_shannon}.}

\bibliography{main}{}
\bibliographystyle{aasjournal}

\end{document}